\documentclass[12pt,preprint]{aastex}
\shorttitle{MARGINALLY COLLISIONLESS CORONAL LOOPS}
\shortauthors{Imada et al.}

\begin{document}

\title{ SELF-ORGANIZATION OF RECONNECTING PLASMAS TO  MARGINAL COLLISIONALITY IN THE SOLAR CORONA}

\author{S. \textsc{Imada},\altaffilmark{1,2} 
and E.G. \textsc{Zweibel} \altaffilmark{3}
}
\altaffiltext{1}{ Institute of Space and Astronautical Science, Japan Aerospace Exploration Agency, 3--1--1 Yoshinodai, Chuo-ku, Sagamihara-shi, Kanagawa 252--5210, Japan}
\altaffiltext{2}{ Cuurent address: National Astronomical Observatory of Japan, 2-21-1 Osawa, Mitaka-shi, Tokyo 181-8588, Japan}
\altaffiltext{3}{Departments of Physics and Astronomy and Center for Magnetic Self-Organization, University of Wisconsin-Madison, 475 N Charter Street, Madison, WI 53706, USA}

\begin{abstract}
We explore the suggestions by Uzdensky (2007) and Cassak et al. (2008) that coronal loops heated by magnetic reconnection should self-organize to a state of marginal collisionality.
We discuss their model of  coronal loop dynamics with a
 one-dimensional hydrodynamic calculation. We assume that many current sheets are present, with a distribution of thicknesses, but that only current sheets thinner than the ion skin depth can rapidly reconnect. This assumption naturally causes a density dependent heating rate which is actively 
regulated by the plasma. We report 9 numerical simulation results of coronal loop hydrodynamics in which the absolute values of the heating rates are different but their density dependences are the same. We find two regimes of behavior, depending on the amplitude of the heating rate. In the case that the amplitude of heating is below a threshold value, the loop is in stable equilibrium. Typically the upper and less dense part of coronal loop is collisionlessly heated and conductively cooled. When the amplitude of heating is above the threshold, the conductive flux to the lower atmosphere required to balance collisionless heating drives an evaporative flow which quenches fast reconnection, ultimately cooling and draining the loop until the cycle begins again. The key elements of this cycle are gravity and the density dependence of the heating function. Some additional factors are present, including pressure driven flows from the loop top, which carry a large enthalpy
flux and  play an important role in reducing the density. We find that on average the density of the system is close to the marginally collisionless value.

\end{abstract}


\section{Introduction}
Magnetic reconnection has been discussed as one of the important mechanisms for explosive events in astrophysical plasma, 
because the free energy stored in the magnetic field can be rapidly released to the plasma and converted to 
kinetic energy, thermal energy, and non-thermal particle energy.
Over the last several decades, considerable effort has been devoted toward understanding the energy conversion mechanism and distribution rate among kinetic, thermal, and non-thermal energy during reconnection not only in the solar atmosphere \citep[e.g.,][]{pne} and other astronomical objects but also in the Earth's magnetosphere \citep[e.g.,][]{hon,nag1,nag2,bau,oie,ima,ima4} and the laboratory  \citep[e.g.,][]{baum,ono,yam, ji}.
 
In the solar atmosphere, it is believed that magnetic reconnection is the fundamental energy conversion mechanism of eruptive
flares.  It  may also contribute to coronal heating, which
remains one of the essential and fundamental problems in solar physics.
Over the last several decades, considerable effort has been devoted toward understanding coronal heating, and various processes have been discussed \citep[e.g.,][]{man}. 
One plausible mechanism is the nanoflare heating model \citep[e.g.,][]{par1, par2, aly}. 
The essential idea is that the slow convection-driven random motion of the photospheric footpoints of the coronal magnetic field drags the field into complex patterns which
leads to the buildup of many current sheets.
These current sheets cause ubiquitous small scale reconnection events in the corona, releasing magnetic energy to the coronal plasma. 
To clarify the coronal heating mechanism, and identify the conditions and locations where heating takes place, many observational studies have been performed \citep[e.g.,][]{lin,shi,dos}.
Numerical modeling of coronal heating \citep[e.g.,][]{mcc1,mcc2,kli,kli1,war,war2} has also been carried out to study plasma dynamics in coronal loops with different heating models, for example uniform, localized, steady, and non-steady heating. 
Much of this work was reviewed recently by \citet{reale_10}.
 
Reconnection can be of the slow magnetohydrodynamic (MHD) Sweet-Parker type \citep{swe,par} or the fast collisionless type, in which Hall effects are important \citep{son,ter,bir,rog,zwe}. 
Therefore, another important issue for the role of reconnection in astrophysics  is when and where fast magnetic reconnection occurs.
While considerable progress has been made over the past decade, the key question has not yet been answered clearly.
One plausible answer, for which there is experimental evidence, is related to the collisionality of plasma and/or the  current sheet thickness. 
Specifically, it appears that if the reconnection layer thickness predicted by Sweet-Parker theory is less than the ion skin depth $\delta_i\equiv c/\sqrt{4\pi n_ie^2/m_i}$,
reconnection is collisionless; otherwise, it is MHD. This criterion can be written in terms of $\delta_i$, the current sheet length $L_c$, electron
cyclotron frequency $\omega_{ce}$, and electron collision time $\tau_e$ as $L_c/\delta_i <\omega_{ce}\tau_e$ for collisionless reconnection \citep{zwe}. Since $\delta_i\propto n^{-1/2}$
and $\tau_e\propto n^{-1}$, the criterion for collsionless reconnection is easier to satisfy in a lower density plasma, all other
things being equal.

Recently, \citet{uzd} and \citet{cas} proposed independently that coronal loops heated by reconnection could self-organize to a state of marginal collisionality.
Although the outcomes are the same, the arguments presented in the two papers are different. The basic  idea in \citet{uzd} is that if the plasma is near the marginal state and becomes less collisional (hotter, less dense, or both) the heating rate will increase. This drives the temperature up further, increasing the conductive heat flux to the lower atmosphere. The increased heat flux is compensated by an upward evaporative flow, which increases the density in the loop and lowers the heating rate. Conversely, an increase in collisionality leads to cooling and to reduced density as the material settles due to gravity. Once the density is lower, the heating rate increases again. \citet{cas} also invoke a drastic increase in the reconnection rate with decreasing collisionality, but their model depends on how the collisionality parameter depends on temperature and magnetic field, and how these quantities evolve as current sheets form and begin to slowly reconnect. Since our model gives  a statistical description of reconnection, contains few details about how reconnection layers form and how they are structured, and requires gravitational
stratification, it is more closely related to \citet{uzd} than to \citet{cas}.

Similar considerations might hold if fast reconnection is due to onset of the plasmoid instability \citep{lou, hua}. The instability sets in at a critical value of the Lundquist number $S$
of order $4\times10^4$; since $S=\omega_{ce}\tau_e L/\delta_i\propto T^{3/2}/n^{1/2}$, reducing the density could also trigger fast reconnection. However, the plasmoid instability has not yet
been explored in line tied systems such as the solar corona, and we defer further discussion to future work.

In this paper, we study coronal loop dynamics with a heating function which increases sharply with decreasing density, as expected for heating by collisionless magnetic 
reconnection heating. In \S\ref{model} we describe the basic model. In \S\ref{equilibrium} we discuss the nature of the equilibrium solutions. In \S\ref{results} we
present three representative coronal loop models with different amplitudes of the collisionless heating rate. We find stable equilibrium states for low amplitudes and
periodic oscillations at higher amplitudes. We also present a parameter study which shows trends in various quantities with heating rate amplitude. In \S\ref{summary} we summarize and discuss
the results, and
mention other possible applications.

\section{Model}\label{model}
\subsection{Basic Equations}\label{BE}
We consider a single magnetic loop with an arch-like configuration.
The loop has a fixed semi circular shape with a constant cross section, with half-length $L= 26$ Mm.
The loop is taken to have an infinitely strong magnetic field, so that the plasma moves and the heat flows freely along the loop while energy and mass transport across loop 
are strongly inhibited.
Assuming symmetry about the loop top and a fully ionized atmosphere, we calculate the dynamics in only half of the loop using a 1D hydrodynamic (HD) code.
For simplicity, we take  the ions to consist of only protons, though other elements are included in evaluating radiative losses.
We use a single-fluid description, i.e., electrons and ions have the same temperatures and bulk velocities \citep[e.g.,][]{hor,shim}.

The equations of mass, momentum, and energy conservation in Eulerian form are
\begin{equation}
\frac{\partial \rho}{\partial t} + \frac{\partial}{\partial x}\left(\rho V_x \right)=0,
\end{equation}
\begin{equation}
\frac{\partial}{\partial t}\left(\rho V_x \right) + \frac{\partial}{\partial x}\left(\rho V_x^2 +p \right) =-\rho g_{\parallel} ,
\end{equation}
\begin{equation}\label{energy}
\frac{\partial}{\partial t}\left(\frac{p}{\gamma-1} + \frac{1}{2}\rho V_x^2\right)  + \frac{\partial}{\partial x}\left[\left(\frac{\gamma}{\gamma-1}p+\frac{1}{2}\rho V_x^2\right) V_x - \kappa_{\parallel} \frac{\partial T}{\partial x} \right]=-\rho g_{\parallel} V_x + H -R,
\end{equation}
\begin{equation}
p=\frac{k_B}{m}\rho T,
\end{equation}
\begin{equation}\label{g}
g_{\parallel}=g_0\cos\left[\left(\pi/2\right)x/L\right]
\end{equation}
in cgs units. Here $x$ is the distance along a loop from its base, $\rho$ is the proton mass density, $m$ is the
mean mass per particle ($=m_p/2$),
 $v$ is the fluid velocity, $p$ is the total gas pressure, $T$ is the plasma temperature, $g_{\parallel}$ is the solar gravity along the loop, $g_0$ is the gravity at the 
solar surface ($2.74\times 10^4$ cm s$^{-2}$), $k_B$ is Boltzmann's constant, and $\gamma$ is the ratio of specific heats for an ideal gas, taken to be 5/3. The validity of
the 1D approximation can be quantified by noting that a fractional pressure fluctuation $\delta P/P$ induces a fractional radial perturbation $\delta R/R\sim \beta/4$, where
as usual $\beta$ is the ratio of gas pressure to magnetic pressure. As we will see, while the density and temperature fluctuations in the loop are individually large, they are
anticorrelated in time, so the pressure fluctuations are relatively small, and the changes in radius for a low $\beta$ loop will be truly small.  

Heat conduction along the loop is primarily by electrons; we use  the classical conductivity for a fully ionized hydrogen plasma \citep[][]{spit}:
\begin{equation}
\kappa_{\parallel}=\kappa_0 T^{5/2},
\end{equation}
where $\kappa_{\parallel}$ is $9.0 \times10^{-7}$ erg s$^{-1}$ K$^{-1}$ cm$^{-1}$.

The radiative loss rate is denoted by $R$, and is given by
\begin{equation}
R=\rho^2 \lambda_\rho(\rho) \Lambda(T),
\end{equation}
where $\lambda_\rho(\rho)$, $\Lambda(T)$ represent the effect of optical thickness on the efficiency of radiative cooling and the radiative energy loss function,
respectively.
We take $\lambda_\rho(\rho) = \rho_{cl}/\rho \tanh (\rho / \rho_{cl})$, and $\rho_{cl}= m \times 10^{12}$ g cm$^{-3}$ ($m$: the mean mass per particle).
Thus, radiative cooling is strongly suppressed below the transition region, where the atmosphere is optically thick. Our treatment is obviously approximate, but spares us the complexity of the full radiative
transfer problem for the lower atmosphere. 
The steady state radiative loss function, $\Lambda (T)$, in the solar atmosphere has been calculated by many authors \citep[e.g.,][]{tuc, rosner_78}, and their results vary by a factor of 10 or less.
A compilation of various calculations is discussed in \cite{asc}. All the models assume the plasma is collisionally ionized and optically thin. The differences in the various calculations are mainly from different assumptions about
 the elemental abundance in the solar atmosphere. Further, time-dependent ionization can also affect the value of the radiative loss function \citep[e.g.,][]{bra2, ima5}. In the interests of simplicity and to avoid detailed discussion of the radiative loss function, we use an analytical expression, which reproduces past calculations,  for $\Lambda$
\begin{equation}\label{lambda}
\Lambda(T)=\Lambda_010^{\Theta \left(T\right)},
\end{equation}
where $\theta\equiv\log_{10}(T/T_{cl})$, and
\begin{equation}
\Theta(\theta)=0.4\theta-3+6/(\exp(1.5\theta+0.08))+\exp(-2(\theta+0.08))),
\end{equation}
with
$T_{cl}$ taken to be $2 \times 10^5$ K.
Figure 1 shows the radiative loss function of our formula (solid line) and calculated from the CHIANTI atomic database (dashed line).
For coronal plasma, our formula well reproduces the radiative loss function calculated by CHIANTI.
The radiative loss function is underestimated below the transition region ($T < 10^5$ K). Our study is insensitive to this
discrepancy, because of our assumption that radiative cooling does not work below the transition region due to optical thickness.   

The energy input rate per unit volume is $H$, and we divide it into three parts,
\begin{equation}
H(n,T)=H_1(n)+H_2(n,T)+H_3(n,T)
\end{equation}
where $H_1$, $H_2$, $H_3$ represent respectively
 heating by collisionless reconnection, collisional Sweet-Parker reconnection heating, and heating by an unspecified mechanism to maintain the photosphere and 
chromosphere. These heating function will be discussed in the next section.

The calculations described in this paper were performed using the 1D version of the numerical package CANS (Coordinated Astronomical Numerical Software) maintained by Yokoyama et al\footnote{CANS (Coordinated Astronomical Numerical Software) is available online at http://www-space.eps.s.u-tokyo.ac.jp/~yokoyama/etc/cans/. }.
In our calculation, we used 2001 grid points in x. 
Grid spacing below the transition region is set to be 0.01 $h_0 (x < 1.3x_{tr})$, where $h_0$, $x_{tr}$ are the pressure scale height in the chromosphere ($h_{0} = 200$ km) 
and the transition region location ($x_{tr} = 2500$ km), respectively.
In the corona, we use $\Delta x_{i+1}= 1.02 \Delta x_{i}  (x>1.3 x_{tr})$.
We use reflecting boundary conditions at x=0 and L; $\partial \rho/\partial x=0$, $\partial p /\partial x=0$, $V_x=0$.
We impose an upper limit on $\Delta x $ of 0.5 $h_0$. Timestepping is explicit, and set by the CFL condition.

\subsection{The Heating Functions}\label{heating}
Our choices for the three heating functions $H_1$, $H_2$, $H_3$ are physically motivated. However, they are not necessarily universal, or completely accurate.
This is because our knowledge
of the coronal magnetic field and of how magnetic reconnection behaves under coronal conditions is incomplete, and also because we seek a model which is computationally tractable.
Therefore, our results should be considered illustrative.

\subsubsection{Coronal heating model in the collisionless regime}\label{cless}
As we mentioned in the Introduction, one of the possible mechanisms for explaining coronal heating is micro/nano-flare heating. 
In this model there are many current sheets in the solar corona, and small magnetic reconnection events occur within the current sheets. 
This scenario is based on the assumption that the energy source of the coronal heating is convection - driven motion of the photosphere, which has a large amount of energy. In Parker's
original picture \citep{par1,par2} current sheets form because the distorted field cannot adjust to a smooth equilibrium. This point is still not settled, but it is generally agreed
that footpoint motion progressively increases the coronal current density over time, and that the distribution of current is highly intermittent. We associate the intermittency with
current sheets. 
Because the photospheric motions are slow, the spatial scale of initially formed current sheet might be large compared with the dissipation scale. 
It is plausible that many of the current sheets are not dissipated immediately. 
Therefore, we assume that the number of current sheets and frequency
of reconnection events are large enough that the heating rate is statistically almost steady over time.
To discuss the regime transition of magnetic reconnection, we assume that fast reconnection occurs when the thickness of the current sheet become comparable to the ion 
inertial length $\delta_i$.
The thickness of current sheets in solar coronal loops is not well known because measuring the coronal magnetic field is very difficult, and from a theoretical
point of view the
distribution in thickness and formation mechanism are not yet well understood. 
Therefore, we assume that there are many current sheets from small scale ($\sim \delta_i$) to large scale ($\sim$ loop length) in the solar corona, and the dissipation scale 
of the current sheets is defined by ion inertial length $\delta_i$ (Figure 2a).
The current sheet distribution should have a cutoff around the dissipation scale.
Thus, we take the current sheet thickness distribution near the dissipation scale to be
\begin{equation}
f(\delta)=\frac{1}{2}f_0\left(\tanh\left(\frac{\delta-\delta_c}{\lambda}\right)+1\right),
\end{equation}
where $f$, $\delta$, $\delta_c$, $f_0$, and $\lambda$ are the distribution function of the current sheet, the current sheet thickness, 
the critical thickness of the distribution, the distribution function at large scale ($\delta >> \delta_c$), and the transition scale of the distribution.
The critical thickness of the distribution ($\delta_c$) should be around the usual ion inertial length in the solar corona, and we fix its value at
$\delta_c = 720$cm ($n\sim 10^9$ cm$^{-3}$).   
We take the transition scale ($\lambda$) equal to $0.1\delta_c$ ( $72$ cm). 
We assume that these parameters do not change during the calculation.
The current sheet distribution function is shown in Figure 2b.

Current sheets which are thinner than the ion inertial length reconnect rapidly and  cause strong heating. 
We can derive the heating rate by collisionless reconnection as follows;
\begin{equation}\label{h1}
H_1(\delta_i)= \dot{E} \int_0^{\delta_i}f d\delta=H_{c1}\left(\frac{\delta_i}{\delta_c} +\frac{\lambda}{\delta_c} \log \left( \frac{\cosh \left(\frac{\delta_i-\delta_c}{\lambda}\right)}{\cosh\left(\frac{\delta_c}{\lambda}\right)}\right)\right),
\end{equation}
where $\dot{E}$, $H_{c1}$ are the energy release rate by fast reconnection in each current sheet, and the heating rate parameter ($=\dot{E}f_0\delta_c/2$),
respectively. Figure 2c shows the heating rate as a function of density in the three cases ($H_{c1} = 3 \times 10^{-2}-10^{-4}$ erg cm$^{-3}$ s$^{-1}$). As a ``reality check"
of this parameter range, we define an effective magnetic dissipation time $\tau_{diss}$ by $H_1=B^2/(8\pi\tau_{diss})=400 (B/100G)^2/\tau_{diss}$ erg cm$^{-3}$ s$^{-1}$. We then see that
the shortest dissipation time $\tau_{diss}$ shown in Figure 2c is about 400s for a 100G magnetic field, and occurs for $H_{cl}=3\times 10^{-2}$ erg cm$^{-3}$ s$^{-1}$ and $n=
10^8$ cm$^{-3}$. However, in later section (\S 4.3) we show that the loop density never falls below 10$^9$ cm$^{-3}$, corresponding to $\tau_{diss}$ an order of magnitude larger. We conclude
that this range of parameters is reasonable for average magnetic loops in the solar corona. However, in \S 4.5 (Parameter survey) we show that much larger heating rates can be
achieved if $H_{cl}$ is as large as $3\times 10^{-1}$ erg cm$^{-3}$ s$^{-1}$. Such a large heating rate demands rapid dissipation of the free energy of a very large background 
field which is highly stressed, which may not be commonly achievable. But we think it is useful when we apply our model to other astrophysical conditions.

\subsubsection{Coronal heating model in the collisional regime}\label{c}
For collisional heating, we assume that Sweet-Parker reconnection is the dominant heating mechanism.
The energy release rate in Sweet-Parker reconnection is proportional to  $V_AS^{-1/2}$ in the case of constant magnetic field.
The Spitzer resistivity  $\eta \propto T^{-3/2}$. 
Thus we assume the heating rate in the collisional regime can be written as
\begin{equation}
H_2(T)=H_{c2} \times \left(\frac{T}{T_c}\right)^{-\frac{3}{4}}\left(\frac{\rho}{\rho_c}\right)^{-\frac{1}{4}},
\end{equation}
where $H_{c2}$, $T_c$, $\rho_c$ are the parameters for collisional heating, which we take to be
$1.0\times 10^{-5}$ erg cm$^{-3}$ s$^{-1}$, 2 MK, and $m \times 10^9$ g cm$^{-3}$, respectively.

\subsubsection{Chromospheric and lower transition region heating}\label{u}
Generally, the heating of the chromosphere to lower transition region is believed to be larger than that of the corona, because of their high density condition.
These regions have an important role in our study as the mass reservoir or energy consumer of the excess energy in the corona.
Therefore, to produce a robust chromosphere and lower transition region we assume an unspecified heating mechanism of the form
\begin{equation}
H_3(\rho,T)=\frac{H_{c3}}{2}\frac{\rho}{\rho_0}\left(\frac{T}{T_0}\right)^{-\frac{3}{4}} \left(1+\tanh \left( \left(\frac{\rho}{\rho_{cl}}-1 \right)/ \lambda_3\right)\right),
\end{equation}
where $H_{c3}$ ,  $\rho_0$, $T_0$ are the heating rate coefficient ($6\times 10^4$ erg cm$^{-3}$ s$^{-1}$), mass density ($m \times 10^{17}$ g cm$^{-3}$), and chromospheric
reference temperature ($10^4$ K). The parameters
$\rho_{cl}$, $\lambda_3$ are taken to be $10^{10}$ and 0.1; this reduces $H_3$ nearly to zero in the corona, as desired.

\section{Equilibrium}\label{equilibrium}

Static solutions of Equation (\ref{energy}) satisfy the thermal equilibrium condition
\begin{equation}\label{te}
\frac{d}{dx}\kappa_{\parallel}\frac{dT}{dx}  = R- H.
\end{equation}
Approximating the conductive term by $\kappa_{\parallel}T/L^2$, we can estimate the relative importance of conduction and radiation in cooling the loop by computing the so-called Field length{\footnote{The name recognizes G.B. Field's influential paper on thermal instability (Field 1965).}}, $L_f$, the value of $L$ for which these terms are equal
\begin{equation}\label{lf}
L_f\equiv\left(\frac{\kappa_{\parallel}T}{R}\right)^{1/2}=9.5\times 10^9\left(\frac{T_6^{7/2}}{n_9^2\Lambda_{-23}}\right)^{1/2}{\rm{cm}},
\end{equation}
where for any quantity $q$ the notation $q_a$ means $q/10^a$. For typical quiet solar corona parameters ($T_6\sim 2$, $n_9\sim 1$), $L_f$ well exceeds our chosen
loop half-length 2.6 Mm. Thus, we expect the loop to be cooled primarily by thermal conduction rather than radiation, a point first emphasized by \citet{rosner_78}.

Since $L < L_f$, we would expect conduction to play
an important role in stabilizing the loop. According to the classical analysis of \citet{field_65}, in the absence of conduction both the radiative loss and collisionless
heating functions should destabilize the medium
 to quasi-isobaric perturbations, those for which $\delta n/n\sim - \delta T/T$. A positive temperature fluctuation $\delta T$ is accompanied
by a negative density fluctuation $\delta n$, which increases the heating rate (see Equation \ref{h1}) and lowers the cooling rate, enabling the perturbation to grow. 
However, since $L/L_f\ll 1$, any perturbation which satisfies periodic boundary conditions and
is symmetric about the loop top should be  strongly damped by conduction. Instead, we will see that conduction is destabilizing, because it drives mass exchange with the lower
atmosphere.

Gravitational stratification of coronal loops is usually neglected, but it is critical for our models, so we estimate it here. Using Equation (\ref{g}) and assuming the loop
is isothermal, we find the density $\rho(x)$ to be
\begin{equation}\label{rho}
\rho(x) = \rho(0)e^{-\frac{2L}{\pi H}\sin{[(\pi/2)(x/L)]}},
\end{equation}
where $H\equiv kT/mg_0$ is the thermal scale height. Inserting numerical values into Equation (\ref{rho}) we see that that the density drop from the loop base to its top is 
$e^{-0.27/T_6}$, or about 15\% for a 2 MK loop. Figure 2c shows that even this small difference in density results in a large change in the collisionless heating rate.

Finally, we comment on the global equilibrium. Integrating Equation (\ref{te}) over half the loop length and assuming that the loop top is a temperature maximum leads to the
result
\begin{equation}\label{global}
\kappa_{\parallel}\frac{dT}{dx}\vert_0 = \int_{0}^{L}(H-R)dx.
\end{equation}
Equation shows that any heating in excess of what is lost to radiation - which, for $L_f/L\gg 1$ is most of the heat - is conducted downward through the loop base. The
lower atmosphere must either radiate this heat away or expand upward. While for modest heat fluxes the adjustment is minor, we expect that for downward fluxes which exceed 
some maximum value, evaporation from the lower atmosphere increases
the mass density in the loop. This increases the radiative cooling rate and decreases the collisionless heating rate. In this sense, a collisionlessly heated loop,
viewed as an integrated system, has
negative specific heat: increasing the collisionless heating coefficient $H_{cl}$ drives mass into the loop, resulting in a cooler, denser state. 

It would appear from this argument that no matter how large the collisionless heating coefficient $H_{cl}$ is taken to be, the loop can find a static equilibrium state.
This global view, however, is incomplete. It ignores the effects of stratification, which leads to strong collisionless heating of the loop top, and to highly dynamical
behavior. This is shown in the next section.

\section{Results }\label{results}
Here we report on 9 
numerical simulations of coronal loop hydrodynamics in which the collisionless heating coefficient
$H_{c1}$ varies from $3\times 10^{-5}$ erg cm$^{-3}$ s$^{-1}$ to $3\times 10^{-1}$  erg cm$^{-3}$ s$^{-1}$. First
we discuss three examples of the calculation in detail, then we discuss the trends revealed by this parameter survey.

\subsection{Case I: Steady coronal loop ($H_{c1}=3\times 10^{-4}$)}\label{I}
First we discuss coronal dynamics with relatively weak collisionless heating ($H_{c1}=3\times 10^{-4}$).
We assume the loop is initially in hydrostatic equilibrium.
We set the temperature along the coronal loop as follows;
\begin{equation}
T(x)=T_0+\frac{1}{2}\left(T_{top}-T_0 \right)\left(\tanh \left(\frac{x-x_{tr}}{0.5h_0}\right)+1 \right),
\end{equation}  
where $T_0$, $T_{top}$, $x_{tr}$, and $h_0$ are 
the temperature in the chromosphere ($10^4$ K), the temperature at the loop top (2 MK), transition region 
location (2500 km), and pressure scale height in the chromosphere (200 km). 
Because our initial condition does not satisfy thermal equilibrium, the plasma dynamically changes its condition after starting the calculation, and eventually
calms down to hydrostatic and thermal equilibrium within a few 100 seconds ($\sim L_f/v_s$).

Figure 3 shows the equilibrium state of this model.
Temperature, density, and velocity profiles are shown in Figures 3a,b,c. The temperature and density show the expected sharp chromosphere - corona transition, and the
expected slight increase in $T$ and decrease in $n$ within the corona. The three components of the heating rate are shown in Figure 3d. The dashed line is $H_3$; it dominates
in the chromosphere and is ignorable in the corona.
 The dotted line is collisional heating, $H_2$. It dominates in the lower corona. Collisionless heating ($H_1$) dominates
in the upper corona, especially at the top. The radiative cooling rate is shown in Figure 3e. In the upper corona it is about 2 orders of magnitude less than the
heating rate. The divergence of the conductive heat flux, however, which is shown in Figure 3f, approximately balances the heating rate, as suggested by our computation of the Field length (Equation \ref{lf}). The solid blue curve in Figure 3f shows the regions which are conductively cooled while the dotted red curve shows the regions which are
conductively heated.

\subsection{Case II: Micro-flaring coronal loop ($H_{c1}=3\times 10^{-3}$)}\label{II}
The coronal loop has completely different dynamics if the collisionless heating coefficient $H_{c1}$  is increased by an order of magnitude from Case I, to $3\times 10^{-3}$.
In this calculation, we used the hydrostatic and thermal equilibrium condition in the case of relatively weak collisionless heating ($H_{c1}=3\times 10^{-4}$, Figure 3) for the initial condition. 
Figure 4 shows time series of the result during first 1060 seconds.
From the top, temperature, density, velocity, heating rate, and radiative cooling and thermal conduction are shown.
The figure format for the heating rate is the same as Figure 3d.
Radiative cooling is represented by the solid line in the bottom panel of Figure 4.
The heating and cooling by thermal conduction are shown by dotted and dashed lines, respectively. In order to understand what is going on, compare the adjacent columns at 120
and 660 s. At the earlier time the loop top is hotter (several MK) and less dense, consistent with the strong collisionless heating shown in the fourth panel in the column. 
At the earlier time there is a subsonic upflow ($V>0$), corresponding to evaporation. As the bottom panel shows, the upflow region is conductively heated. At the later time the
loop in cooler and denser. Collisionless heating is weak, and there is a slow downflow due to hydrostatic settling. By 1060 s, the state of the loop has returned to what it
was at 120 s. The range of behaviors is summarized in Figure 5, which shows the spatial profiles of temperature, velocity, density, and collisionless heating rate.

We have continued our calculation until 20000 seconds and found that this cyclical behavior is regular and repeatable.  
Figure 6 shows the long time variation of several coronal loop properties.
From the top, loop top density, loop top temperature, collisionless heating rate integrated along the loop, and total X-ray emission ($1-8 {\rm \AA}$) are shown, respectively.
Total X-ray emissions are normalized by the total X-ray emission in the initial condition.  
We can clearly see $\sim 900$ second oscillation in Figure 6. 
Roughly speaking, collisionless heating is enhanced under low density conditions, and is suppressed under high density conditions. The secondary peak of loop top density in Figure 6 is caused by the evaporation flow which is once reflected at the loop top and reflected back from the loop base to the top.
Because the soft x-ray emissivity varies 
over several orders of magnitude during the cycle, we call the bursts of peak emission microflares. Of course, we have not included enough reconnection physics to show
whether other flare signatures such as hard X-rays from nonthermal electrons would be produced. But we can say that the peak pressure, which 
remains within an order of magnitude of the mean pressure, is probably not enough to disrupt a low $\beta$ loop.

\subsection{Case III: Flaring coronal loop ($H_{c1}=3\times 10^{-2}$)}\label{III}

As the third example, we consider even stronger collisionless heating
($H_{c1}=3\times 10^{-2}$). As in the previous case, we used
the hydrostatic and thermal equilibrium condition in the case of relatively weak collisionless heating ($H_{c1}=3\times 10^{-4}$, Figure 3) for the initial condition.
 
Figure 7 shows the time series of the result during the first 1460 seconds.
The figure format is the same as Figure 4.
Plasma is intensely heated in the higher corona. The loop top becomes overpressured, which drives downflow. The enthalpy transport associated with the downflow,
together with heat carried by thermal conduction, already reaches the chromosphere and causes chromospheric evaporation at $t=20$ seconds. By 60 seconds, the flow is
supersonic    
(a few hundred km sec$^{-1}$) and appears to form a shock. Due to mass transport into the loop,
the heating rate has already dramatically decreased at t=300 seconds. 
Thus, the temperature in loop decreases, and coronal plasma falls down along the loop.
In this case, the downflow caused by the cooling of coronal plasma is relatively large (although not as large as the upflow), and we can clearly see a few tens km sec$^{-1}$ downflows at $t=1300$ sec. 
At that moment, the coronal loop is already colder than the initial condition. 
Afterwards, the density in the higher corona become very low, and the collisionless heating rate increases again.
At 1460 seconds, chromospheric evaporation is occurring again. 
Note that we can still observe the downflow and strong heating at the loop top even during chromospheric evaporation. 
We can easily compare the plasma parameters in different times in Figure 8 and confirm the scenario which is discussed above.

We have continued our calculation until 20000 seconds and found that the flare cycle faithfully repeats.
Figure 9 shows the long time variation of coronal loop.
The figure format is the same as Figure 6.
We can clearly see the $\sim 1500$ second oscillation in Figure 9. This is longer than the $\sim 1000$ s period observed in the previous micro-flare study, Case II.
The peak of the oscillation is sharper than that of micro-flare result, and the asymmetry between density and temperature buildup and decay is more pronounced. This is
because the buildup is driven by collisionless heating, which at its peak is ten times stronger than in Case II (compare Figures 6 and 9). The cooling time is longer,
which accounts for the longer period.

\subsection{Cycle Mechanism}\label{CM}

Another way to understand the cyclic behavior is to compare plasma parameters at different locations as they evolve in time. This is done in Figure 10 for the flaring loop discussed in \S\ref{III}.
Figure 10 shows the temporal variation of temperature, density, velocity, collisionless heating rate, and thermal conduction from 1000 to 2000 seconds.
The solid and dashed lines show the temporal variation at $x= 4.5$Mm (base of loop) and $x=20$Mm (near the loop top).
Around $t=1000$ seconds, the collisionless heating rate is relatively low even near the loop top because of the high density ($\sim 2\times10^{9}$ cm$^{-3}$). 
Thus, the plasma in the coronal loop falls down along the magnetic field.
This downflow causes a reduction of density at the loop top, which naturally increases the collisionless heating rate. By $t\sim 1300$s, the pressure at the loop top is higher than at the base,
because the heating rate at the loop base is still very low.
This pressure gradient causes downflow along the loop and further reduction of the density at the loop top. By $t\sim 1400$s, the collisionless heating rate is rapidly increasing.
Almost simultaneously,  the thermal conductive and enthalpy fluxes reach the chromosphere and cause evaporation.
The evaporation  increases the density in the corona, but the
evaporating material does not reach the loop top for a few tens of seconds.
Thus, even during the time that chromospheric evaporation is occurring, the collisionless heating at the loop top is increasing. 
Around $t=1450$ seconds, the evaporating plasma reaches the loop top, and the density at the loop top increases dramatically.
Then, the heating rate drops significantly. The plasma falls down reducing its density until switching strong collisionless heating on.
Figure 11 shows the schematic illustration of this recurrent flaring scenario.
For the most part, our scenario in Figure 11 is consistent with the scenario discussed in \cite{uzd}.
However, in our scenario the coronal plasma cools down through not radiation but conduction. 
The radiative cooling effectively works only at the transition region. 
Figure 7 ($t=$300 sec) clearly shows that conductive cooling is dominant in the solar corona.
Further, the pressure driven down flow from the loop top, which is clearly seen in Figure 10, plays an important role in reducing the density of the loop top in our scenario, and can carry
a large enthalpy flux.

\subsection{Parameter Survey}\label{P}
 
Plots of the trends in parameters for all nine models as functions of the collsionless heating coefficient $H_{cl}$ are shown in Figure 12.
The crosses, squares, and diamonds show the maximum, average, and minimum, respectively. Filled (overlapped) symbols represent models which settled down to a static equilibrium; other
solutions are oscillatory.
The black and red dashed lines show the fitting result by $H_{c1}^\Gamma$, where $\Gamma$ represents the power law index.
In the top panel of Figure 12, $\Gamma$ is 1 and 0.4 for red (maximum) and black (average) line, respectively.
For the oscillation period, $\Gamma$ is 0.25.
For the loop top density,  $\Gamma$ is 0.5 and 0.3 for red (maximum) and black (average) line, respectively.
In the bottom panel, $\Gamma$ is 0.3 and 0.08 for red (maximum) and black (average) line, respectively.
In general, the amplitude of the oscillations increases with increasing $H_{cl}$, as do the peak temperatures.

The maximum values of total collisionless heating (top cross in Figure 12) has a linear dependence on $H_{cl} (> 3 \times 10^{-3})$.
This is because the minimum of loop top density is roughly constant in each case. At these large values of the heating rate coefficient, reducing $n$ much below 10$^{9}$ cm$^{-3}$ drastically
increases the heat and enthalpy fluxes to the lower atmosphere, driving strong upflows which prevent the density from dropping further.
The power law index of the mean values of total collisionless heating (top square in Figure 12) is less than 1, because in most of the time the collisionless heating is off during the oscillation. 
Note that the two models with the largest values of $H_{cl}$ depart from the trends followed by the other models. Their minimum loop top densities are higher than
predicted by the power law fits while their maximum collisionless heating rates and maximum temperatures are lower (but still higher than in the other models). 
These effects together lead to shorter cooling times than predicted by the power law scaling. Since, as we saw from the asymmetric times profiles in Figures 6 and 9, the
cycle periods are set by the decay time, not the rise time, this leads to slightly shorter oscillation periods for the highest $H_{cl}$ values than otherwise expected.
The radiative energy loss in the higher corona is not negligible for the highest $H_{cl}$ ($> 3 \times 10^{-2}$ erg cm$^{-3}$ s$^{-1}$) values.

\section{Summary and Discussion}\label{summary}

This paper was motivated by the suggestion that a gravitationally stratified plasma heated by magnetic reconnection hovers near marginal collisionality \citep{uzd,cas}. 
The basic idea is sketched in Figure 11. Small increases in density reduce the heating rate. The plasma then cools and settles, increasing the heating rate again. If the heat flux is large enough it drives evaporation from the lower atmosphere, which increases the loop density and lowers the heating rate.
The main difference between this model
and past coronal heating studies is that the plasma actively decides its heating rate. On average, the system maintains a marginally collisionless density. 

In order to explore the model, we
studied coronal loop hydrodynamics with a density sensitive heating function based on
the assumption that many current sheets are present, with a distribution of thicknesses, but that only current sheets thinner than the ion skin depth $\delta_i$ can rapidly reconnect (\S\ref{cless}). 
While the specific form of our adopted heating function is unlikely to be completely realistic, it does have the dependence on density invoked in the models of \citet{uzd} and \citet{cas}. We also
commented in the Introduction that similar behavior might be found if the reconnection rate is mediated by breakup of the current sheets into plasmoids \citep{lou,hua}. 
We adjusted the magnitude of the heating rate
through the parameter $H_{cl}$, but kept its shape fixed for this study. This heating function, balanced by a standard optically thin radiative cooling function, results in thermal equilibrium which is locally unstable to isobaric perturbations. In the absence of a chromospheric mass reservoir, it would be stabilized by thermal conduction for loop lengths of interest. However, mass
exchange between the chromosphere and corona leads to highly nonlinear oscillations (\S\ref{equilibrium}).

We found  two regimes of behavior, depending on the value of $H_{cl}$. When $H_{cl}$ is below a threshold value, the loop is in stable equilibrium in which typically the upper, less dense portion is collisionlessly heated and conductively cooled while the lower portion is heated by other mechanisms, including conduction (\S\ref{I}). 
When $H_{cl}$ is above the threshold, the conductive flux to the lower atmosphere required to balance collsionless heating drives an evaporative flow which quenches fast reconnection, ultimately cooling and draining the loop until the cycle begins again (\S \ref{II},\ref{III}).
The key elements of this cycle are gravity and the density dependence of the 
heating function, as predicted by \citet{uzd} and \citet{cas}. Some additional factors are present, including large enthalpy fluxes and pressure driven flows from the
loop top, which play an important role in reducing the density. The amplitude  of the cycle can be so large that the soft X-ray emissivity of the loop varies by as much as 8 orders of magnitude over 
$\sim 20$m, tempting us to identify these events with flares (\S\ref{CM}). 

In \S\ref{BE}, we quantified the transverse expansion of the loop
due to overpressure and argued that it could be neglected for low $\beta$ loops, particularly because the temperature and density fluctuations are anticorrelated. Had we included transverse
expansion, we might have found that it reduces the enthalpy flux to the lower atmosphere. However, transverse expansion also communicates the thermal cycle to other fieldlines, which might enhance
the overall effect. Exploration of these possibilities will have to await more realistic 3D modeling.

Let us compare our results with recent observational and numerical studies of coronal loops.
There are typically two kinds of loops in the solar corona,
"hot loops" ($>$2MK) and "warm loops" ($\sim$1MK).
The hot loops, which were observed in soft X-rays, seem to be in static equilibrium \citep[e.g.,][]{rosner_78}.
On the other hand, most warm loops are inconsistent with static equilibrium.
Recent observations revealed intensity fluctuations of warm loop which can be interpreted as a signature that originates from numerous sporadic coronal heating events \citep[e.g.,][]{sak,vek}.
Many numerical studies have tried to reproduce these two kind of coronal loops, with limited success.
\cite{kli2} raised five discrepancies between observations and theoretical models:
1) warm loops are observed to have a much higher density than is expected given the observed temperature and length,
2) warm loops tend to have a temperature profile that is much flatter than expected for static equilibrium,
3) the density of warm loops decreases with height much more slowly than expected for a gravitationally stratified plasma at the measured temperature,
4) most loops do not have small-scale intensity structure,
5) the lifetime of warm loop is 1000-5000 s, though hot loops have a much larger range of lifetime. 
In our result, there are largely two kind of loops.
One is the steady coronal loop, which is discussed in  \S \ref{I}, and the other is recurrent flaring loops in \S \ref{II} and \ref{III}.
Hot loops might correspond to steady coronal loops.
Warm loops could be interpreted as the cooling phase of recurrent flaring loops.
In Figure 10 we can clearly see that the density in the cooling phase of recurrent flaring loops (t$\sim$2000 sec) is relatively high (0.5-1$\times 10^{10}$ cm$^{-3}$).
The temperature profile also seems to be flat from loop base (solid line) to loop top (dashed line).
In our result, hot/warm loops can be interpreted as a consequence of small/large amplitude of density dependent heating function, respectively.
Although our model is not enough to explain all characteristics of hot and warm loops, we suggest that switching 
collisionless heating on and off might be a key to understand some coronal loop characteristics.

Another important subject for discussion is recent observations of microflares.
Microflares are thought to be caused by the interaction between pre-existing coronal loop and emerging magnetic flux, and many observations support the idea.
\cite{shi2} and \cite{kan} studied the relationship between emerging fluxes and microflares statistically.
Both of studies concluded that the half of microflares are associated with some apparent magnetic field activity.
However, the other half of microflares seem to occur without emerging flux.
Thus, some microflares might be occurring spontaneously, and our result might apply to these.
Another important observational result for microflares is that the loop is hard to identify before the microflare.
\cite{nit} discuss the time evolution of 13 microflare events.  They found
the coronal loop is almost invisible before microflare in some events. 
One of the characteristics in our model is that the coronal loop severely reduces its density before the microflare.
Thus the loop might be almost invisible before the microflare, especially at the loop top.
Our model also indicates downflow, especially near the loop base, before the microflare.
These signatures should be studied in detail with future observations.

Finally, we turn to one of the most dramatic phenomena associated with magnetic reconnection: solar flares.
Over the past few decades, many studies have been carried out to understand the physical mechanism of solar flares, and various models have been proposed. 
Nowadays, it is widely accepted that magnetic reconnection is the fundamental energy conversion mechanism of eruptive flares \citep[the so called CSHKP model,][]{car, stu,hir,kop}. 
The CSHKP flare model predicts that magnetic fields are opened up in association with filament eruption to form a current sheet.
Magnetic field lines in the current sheet successively reconnect to form apparently growing flare loops. 
Many typical features expected from the magnetic reconnection model have been confirmed by modern telescopes. 
These include cusp-like structure in X-ray images \citep[e.g.,][]{tsu}, non-thermal electron acceleration \citep[e.g.,][]{mas}, chromospheric evaporation \citep[e.g.,][]{teri}, reconnection inflow and outflows \citep[e.g.,][]{yok,inn}, plasmoid ejection \citep[e.g.,][]{ohy}, and coronal mass ejections (CMEs) \citep[e.g.,][]{sve}.
However, some flares do not show the typical characteristics expected from the CSHKP model. Soft X-ray images of some  
flares show not cusp-shaped loop structure but a compact-loop shape or a simple loop.
In addition, not all flares are associated with CMEs.
There is still a lot of 
discussion of whether the CSHKP model can explain all flares or not, and the possibility that loop flares do not form current sheets on the global scale \citep[e.g.,][]{alf,spi,uch}.

We have identified a mechanism for coronal loops to spontaneously produce flares. In our model the flares recur periodically, but we
don't expect exact periodicity, because presumably the current sheet distribution itself changes with time. Events which stress the coronal field, such as flux emergence or strong photospheric shear
flow, could increase the number or stored energy in current sheets, causing a loop to transition from an equilibrium to a flaring state.
Another assumption in our model is that the coronal loop shape/length does not change with time.
Changing of shape/length of coronal loops can also cause reduction of density and/or drive flows \citep[e.g.,][]{ima2}.
Both of these effects should also affect the dynamical features of coronal plasma.
We have to explore our model in 2D MHD to discuss these effects, and we will study them in the future.

In this paper, we also do not discuss time-dependent ionization effects. Many recent
studies indicate the importance of time-dependent ionization \citep[][]{rea,ima3,ima5, bra} when strong heating and flows are present. 
As we mentioned before, this process affects the magnitude of the radiative cooling rate.
Further, it is very important to account for nonequilibrium ionization when directly comparing numerical simulations with observations.
Detailed comparison between the model and the observation is necessary to clarify whether our scenario really happens or not in the solar corona.
This is also important future work.

Finally, we mention other plasmas to which our results may apply.
The key elements of our scenario are gravity and the density dependence of the heating function. These are generally present in stellar and accretion disk coronae \citep[e.g.,][]{goo}, offering possibilities for
further theoretical investigation and comparison with observations.


%
\acknowledgments 

 This work was partially supported by the Grant-in-Aid for Young Scientist B (24740130), by the Grant-in-Aid for Scientific Research B (23340045), by the JSPS Core-to-Core Program (22001),  by the JSPS fund \#R53 (gInstitutional Program for Young Researcher Overseas Visits,h FY2009-2011) allocated to NAOJ. We also acknowledge partial support by the Center for Magnetic Self Organization (NSF PHY0821899) and the hospitality of University of Wisconsin. We thank the referee for useful comments.

\begin{figure}
\epsscale{1.0}
\plotone{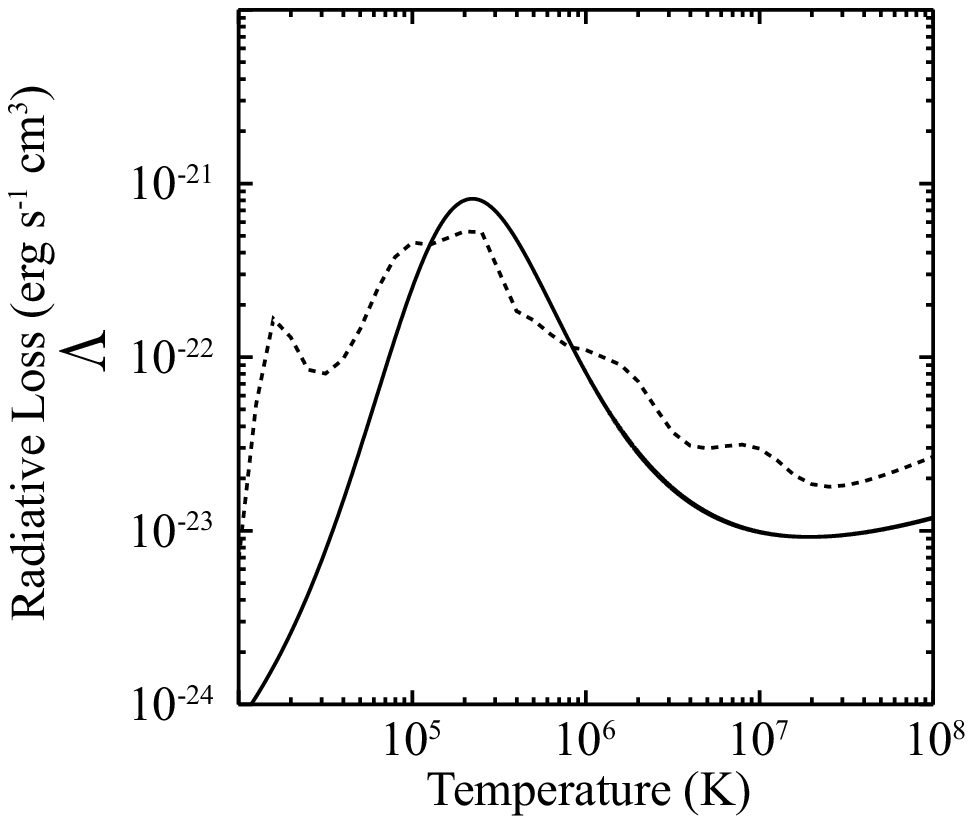}
\caption{Radiative loss function. Solid line shows the function which we used in our study, and dashed line shows the function which was calculated by CHIANTI. Our approximation captures the
correct behavior in the corona; the discrepancies at lower temperatures are less important because of optical depth effects.}
\end{figure}

\begin{figure}
\epsscale{1.0}
\plotone{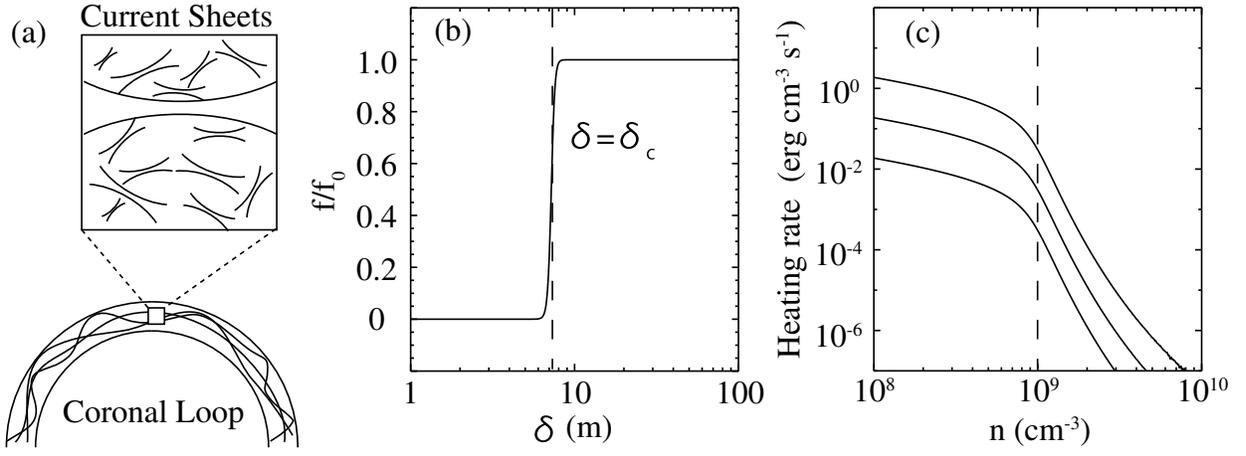}
\caption{Current sheet distribution and Heating rates; a) Schematic illustration of multiple current sheets inside a
coronal loop, b) Current sheet distribution given by Equation (11) as a function of current sheet thickness, c) Collisionless heating rate as a function of density for the three values of $H_{cl}$
used in the models discussed in \S\ref{I} -\ref{III}.}
\end{figure}

\begin{figure}
\epsscale{1.0}
\plotone{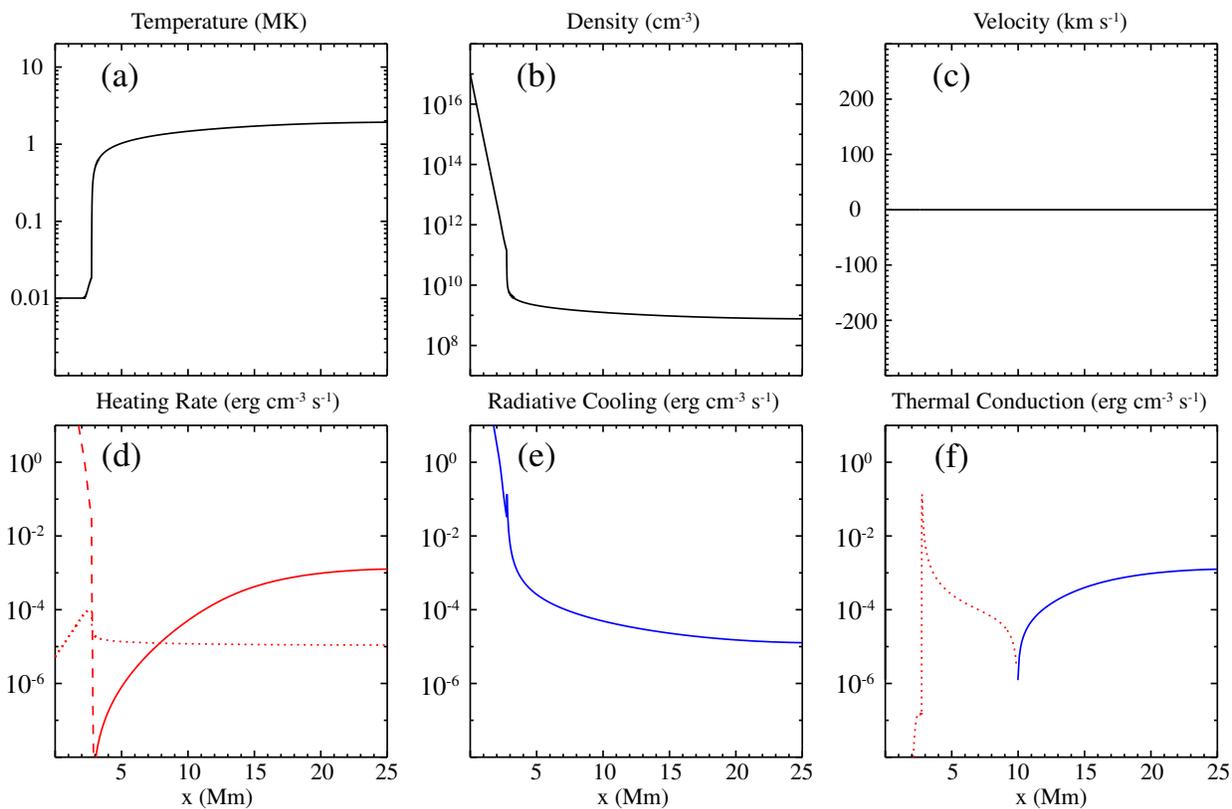}
\caption{ Plasma conditions in state of static equilibrium, achieved for $H_{cl}=3\times 10^{-4}$ erg cm$^{-3}$ s$^{-1}$. In (d), the dashed line represents $H_3$ (chromospheric heating), the
dotted line is $H_2$, collisional reconnection heating, and the solid line is collisionless heating. In (f) the dotted part of the curve represents conductive heating, and the solid part,
conductive cooling.}
\end{figure}

\begin{figure}
\epsscale{1.0}
\plotone{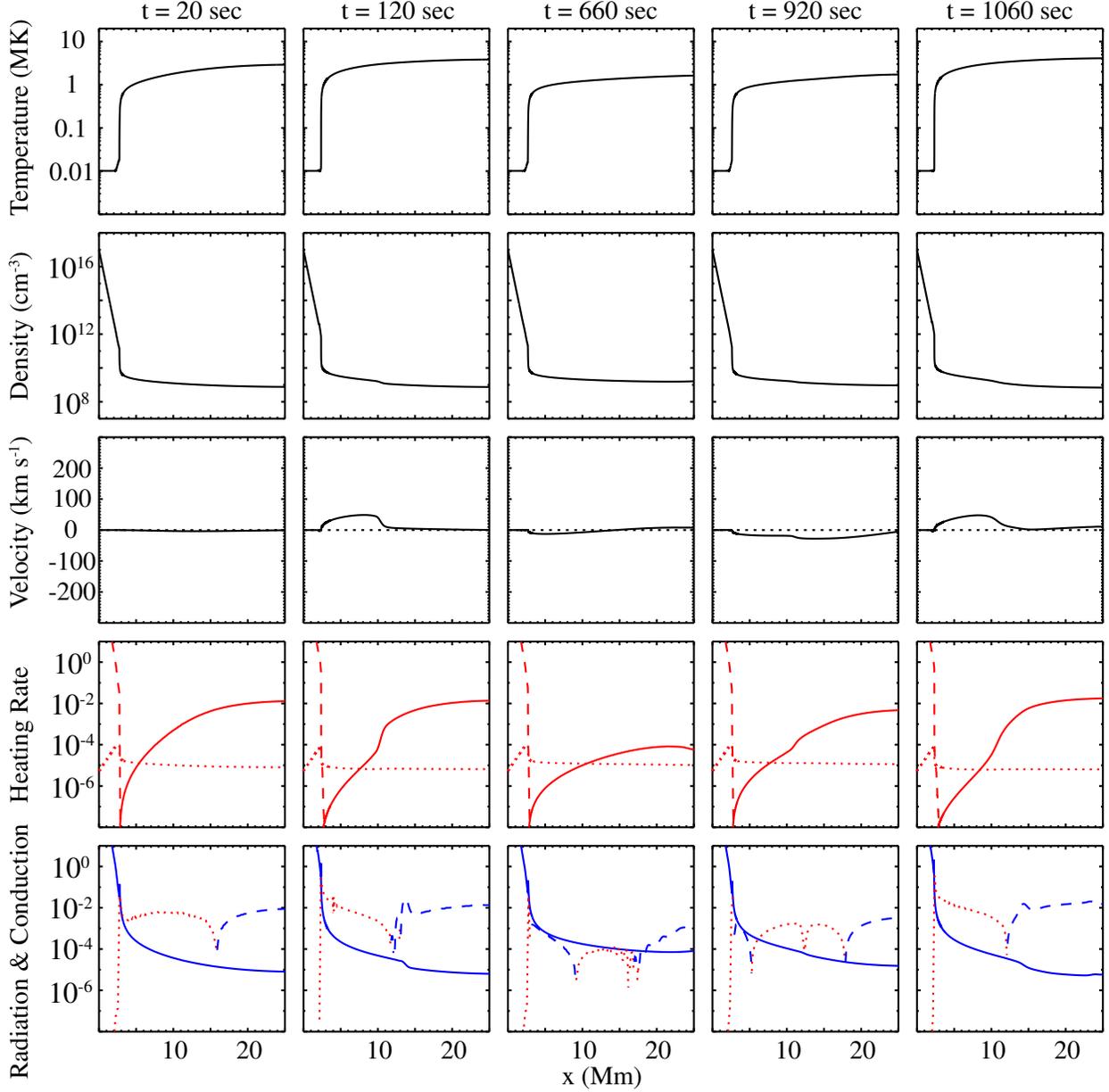}
\caption{ Time series of temperature, density, velocity, heating rate, and radiative cooling \& thermal conduction in the case of micro/nano-flaring coronal loop ($H_{cl}=3\times 10^{-3}$ erg
cm$^{-3}$ s$^{-1}$). In the third row of panels (Velocity), the dotted lines show 0 km s$^{-1}$ to distinguish positive and negative flows. The line styles for the heating rates are the same as in Figure 3d. In the bottom row of panels, the solid line represents radiative cooling, the dashed line is conductive
cooling, and the dotted line conductive heating. The sharpest features conductive heating rates are underresolved.}
\end{figure}

\begin{figure}
\epsscale{0.9}
\plotone{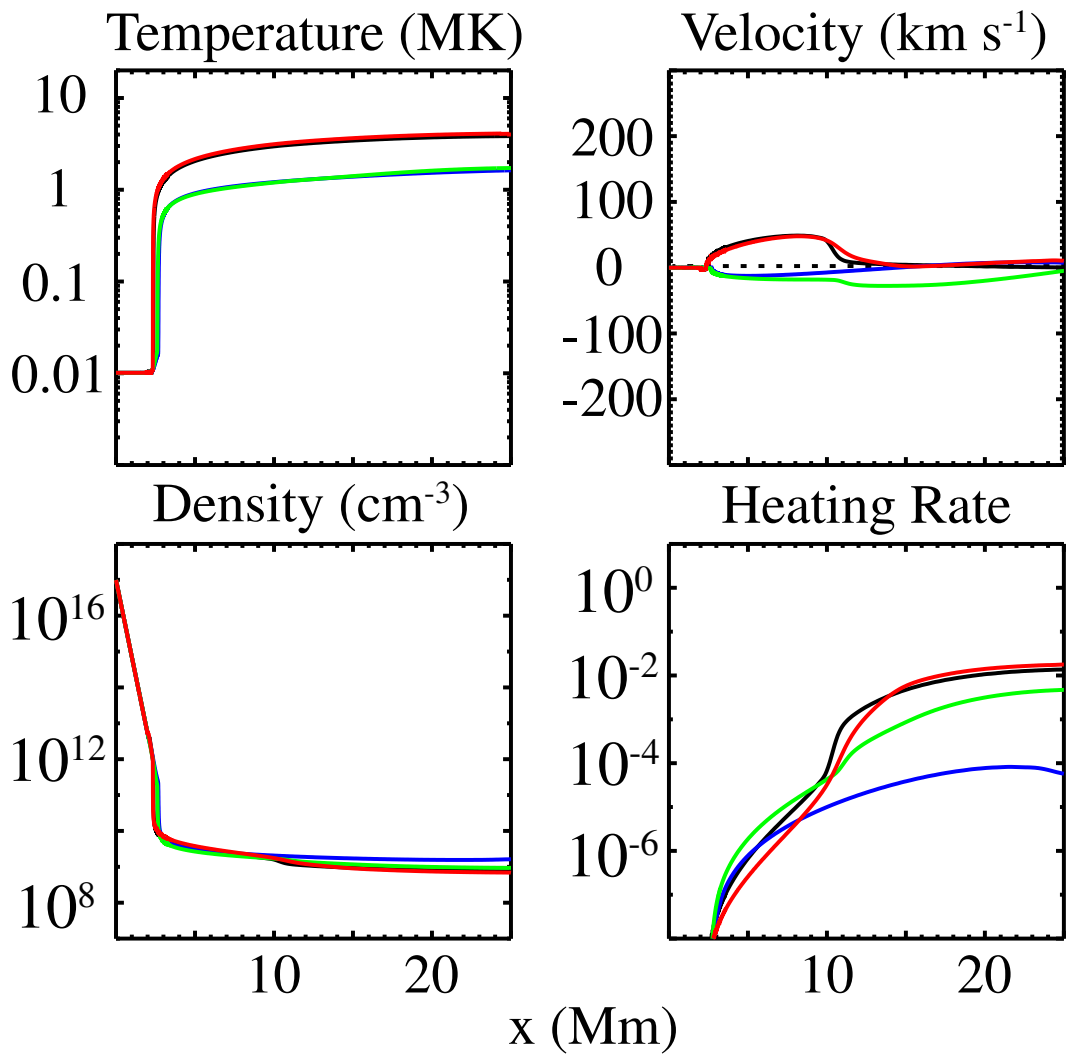}
\caption{ Comparison of plasma conditions among different timing in the case of micro/nano-flaring coronal loop ($H_{cl}=3\times 10^{-3}$ erg cm$^{-3}$ s$^{-1}$. )
Black: t=120s, Blue: t=660s, Green: 920s, Red 1060s. Only collisionless heating
is shown. In the Velocity panels, the dotted lines show 0 km s$^{-1}$ to distinguish positive and negative flows.}
\end{figure}

\begin{figure}
\epsscale{1.0}
\plotone{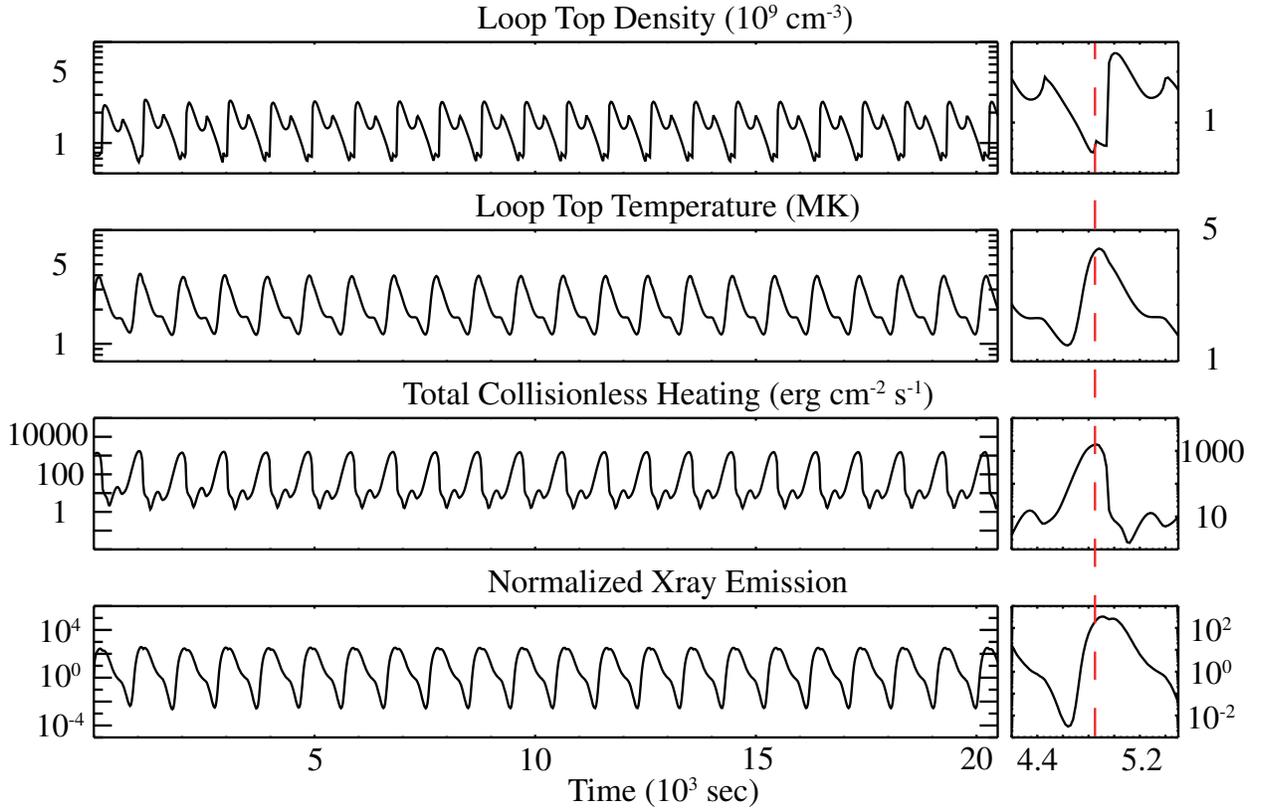}
\caption{ Long term evolution of density, temperature, total collisionless heating, and soft (1 - 8 \AA) X-ray emission, normalized to the emission from the equilibrium loop,
 in the case of the micro/nano-flaring ($H_{cl}=3\times 10^{-3}$ erg cm$^{-3}$ s$^{-1}$) coronal loop. Right;  enlarged figure for one cycle.}
\end{figure}

\begin{figure}
\epsscale{1.0}
\plotone{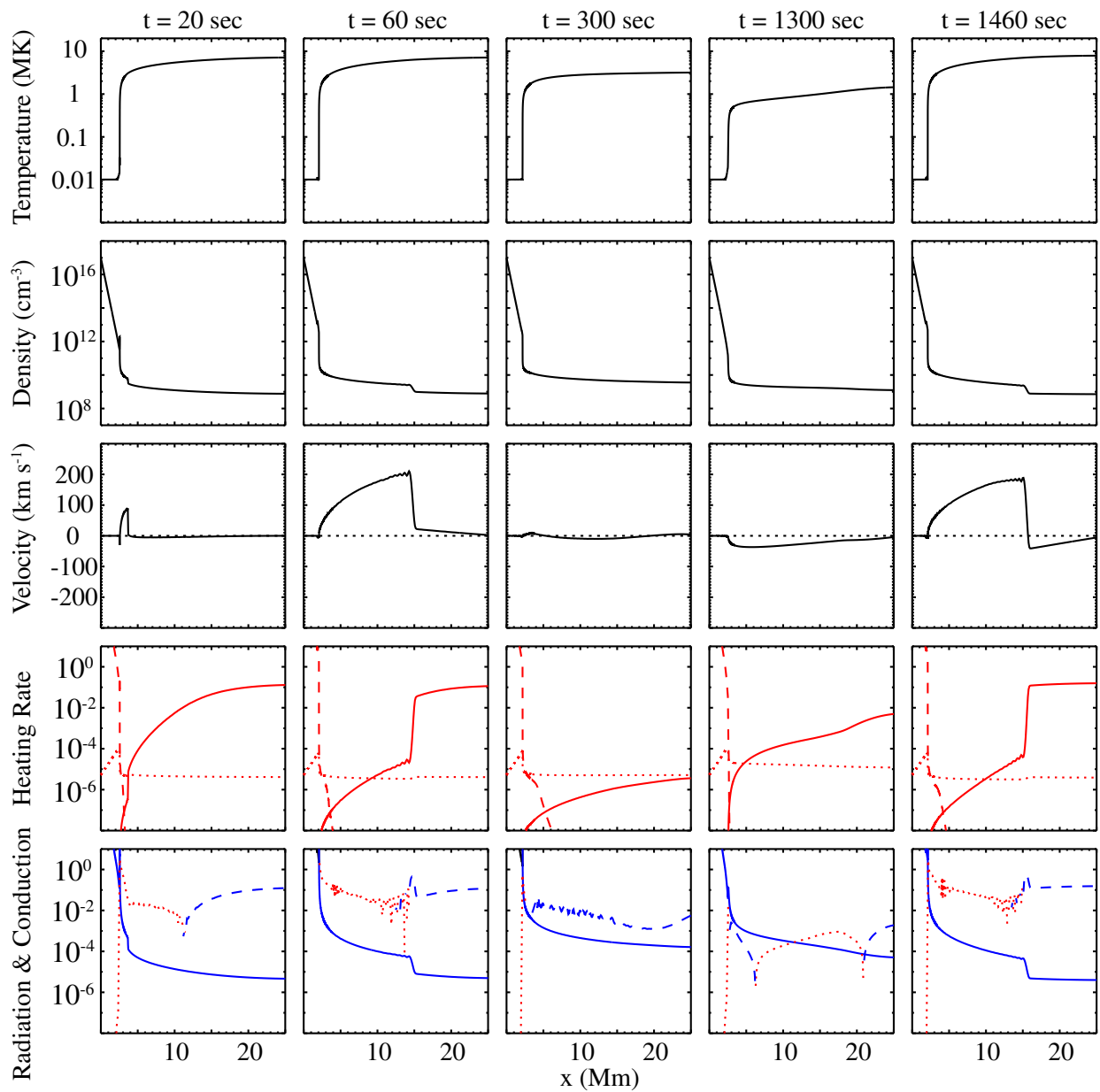}
\caption{Time series of temperature, density, velocity, heating rate, and radiative cooling \& thermal conduction in the case of the flaring coronal loop ($H_{cl}=3\times 10^{-2}$ erg cm$^{-3}$
s$^{-1}$). Format is the same as Figure 4. Some of the sharpest conductive features are underresolved.}
\end{figure}

\begin{figure}
\epsscale{0.9}
\plotone{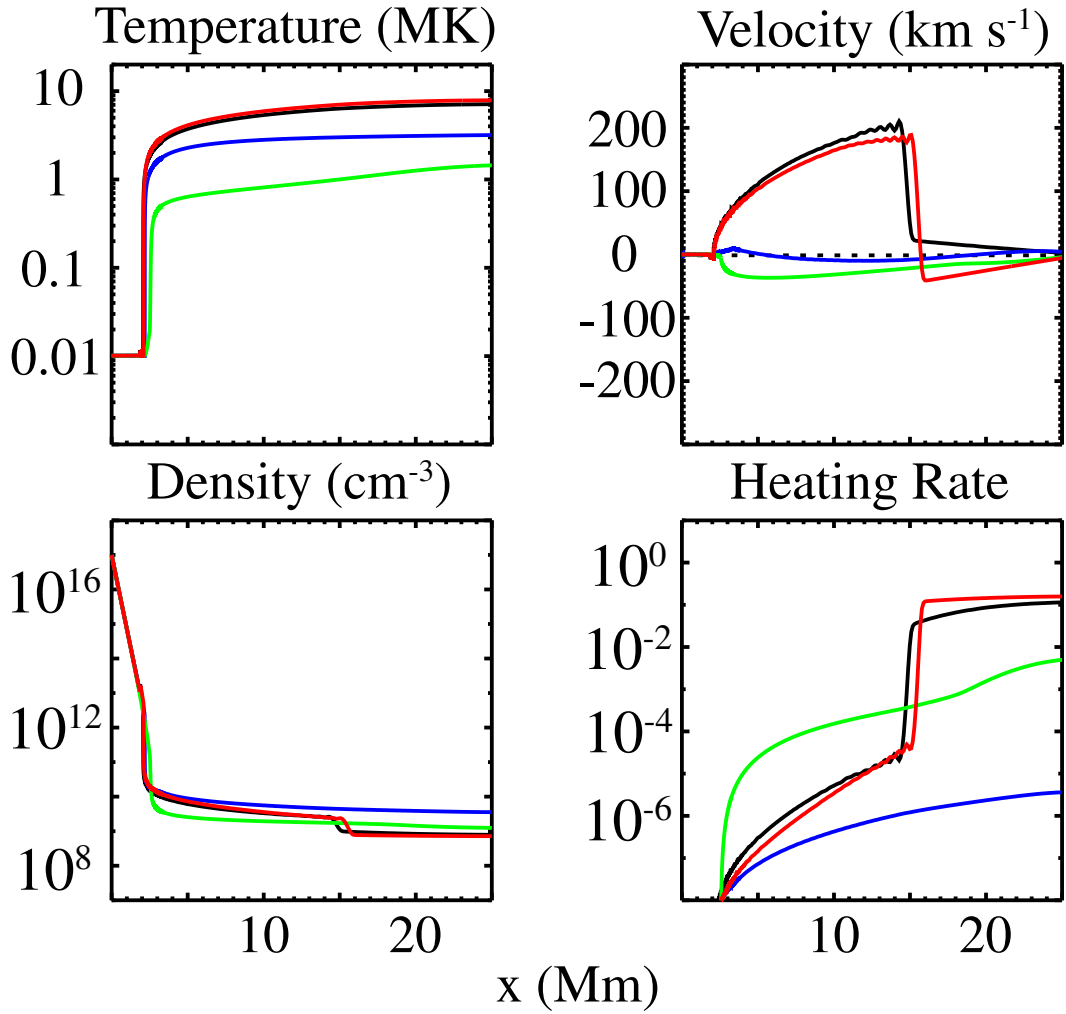}
\caption{ Comparison of plasma conditions among different timing in the case of flaring coronal loop ($H_{cl}=3\times 10^{-2}$ erg cm$^{-3}$ s$^{-1}$). Black: t=60s, Blue: t=300s, Green: 1300s, 
Red 1460s. Format is the same as Figure 5.}
\end{figure}

\begin{figure}
\epsscale{1.0}
\plotone{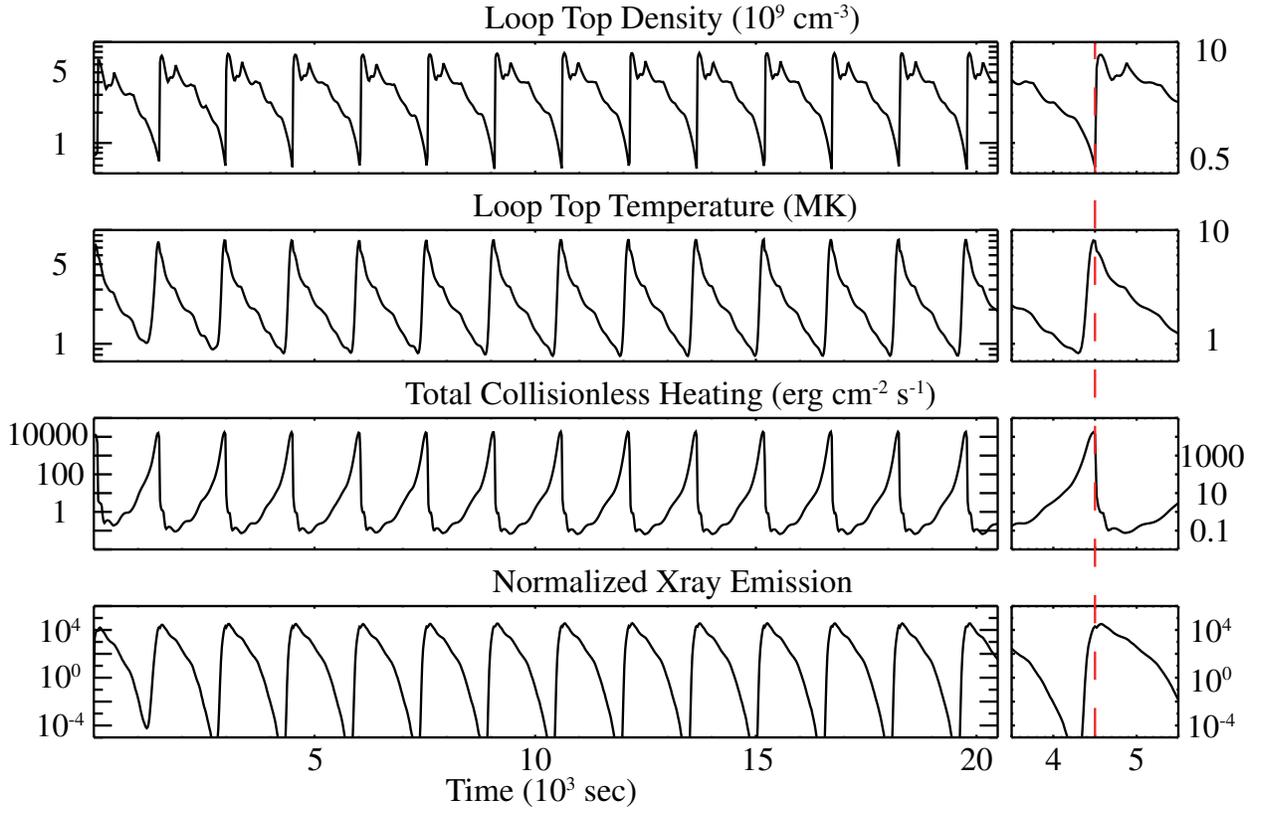}
\caption{ Long term evolution of density, temperature, total collisionless heating, and soft X-ray emission in the case of flaring coronal loop ($H_{cl}=3\times 10^{-2}$ erg cm$^{-3}$ s$^{-1}$). Figure format is the same as Figure 6.}
\end{figure}

\begin{figure}
\epsscale{0.8}
\plotone{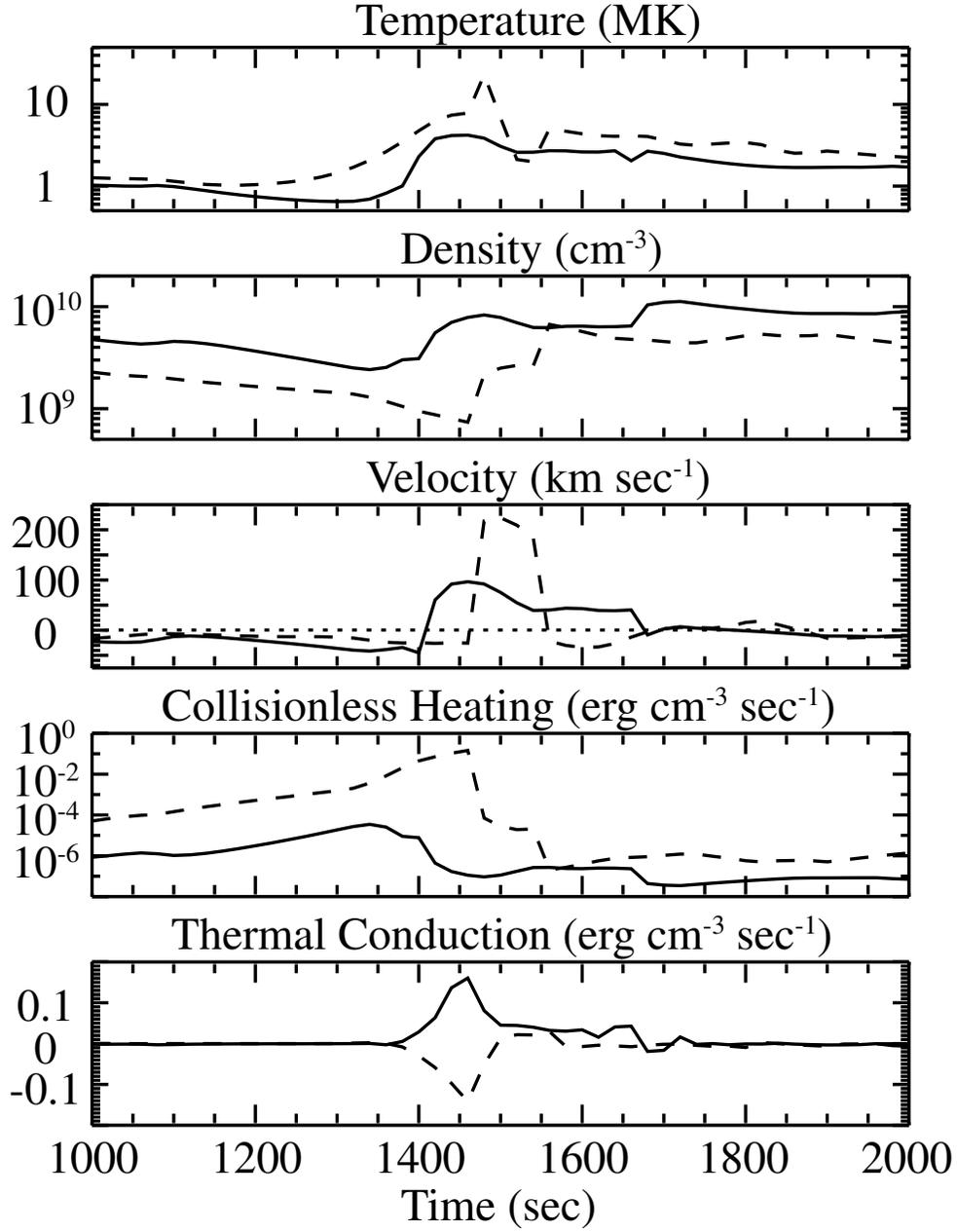}
\caption{ Time evolution of plasma parameters in two different locations of the flaring model ($H_{cl}=3\times 10^{-2}$ erg cm$^{-3}$ s$^{-1}$). The solid line shows the conditions at x=4.5 Mm 
(base of coronal loop), and the dashed line shows  conditions at x=20 Mm (near loop top). In the third row of panels (Velocity), the dotted line shows 0 km s$^{-1}$ to distinguish positive and negative flows.}
\end{figure}

\begin{figure}
\epsscale{0.5}
\plotone{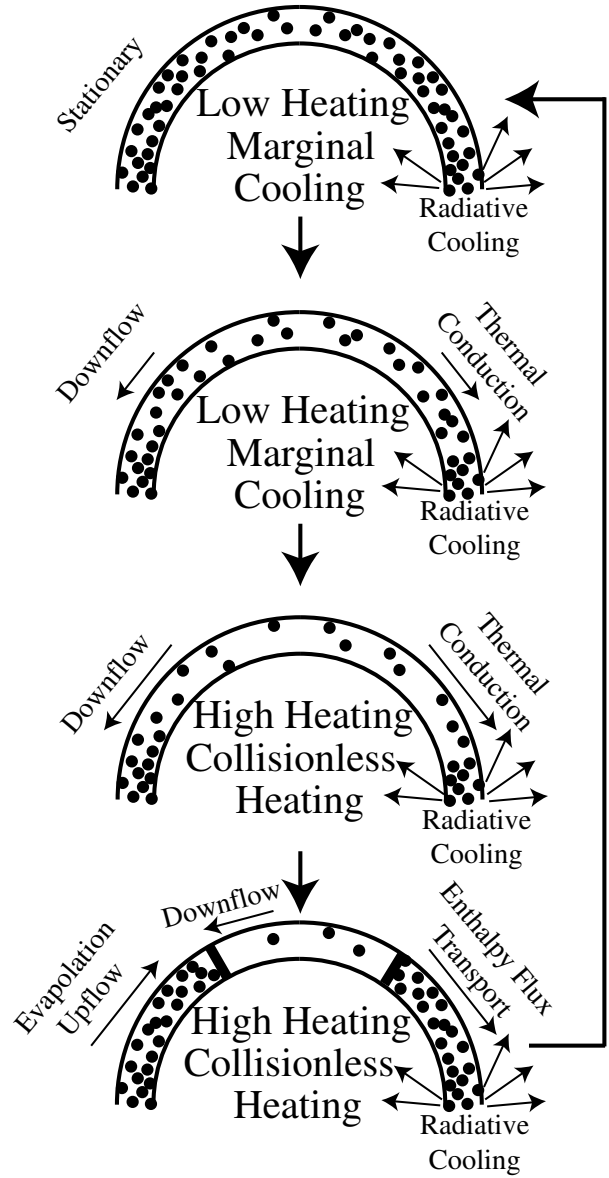}
\caption{ Schematic illustration of the basic idea in our study. A loop which is insufficiently heated undergoes gravitational settling, which lowers the density and increases the heating rate.
The resulting heat fluxes from conduction and pressure driven enthalpy transport drive an evaporative upflow, which quenches the heating and starts the cycle anew.}
\end{figure}

\begin{figure}
\epsscale{0.33}
\plotone{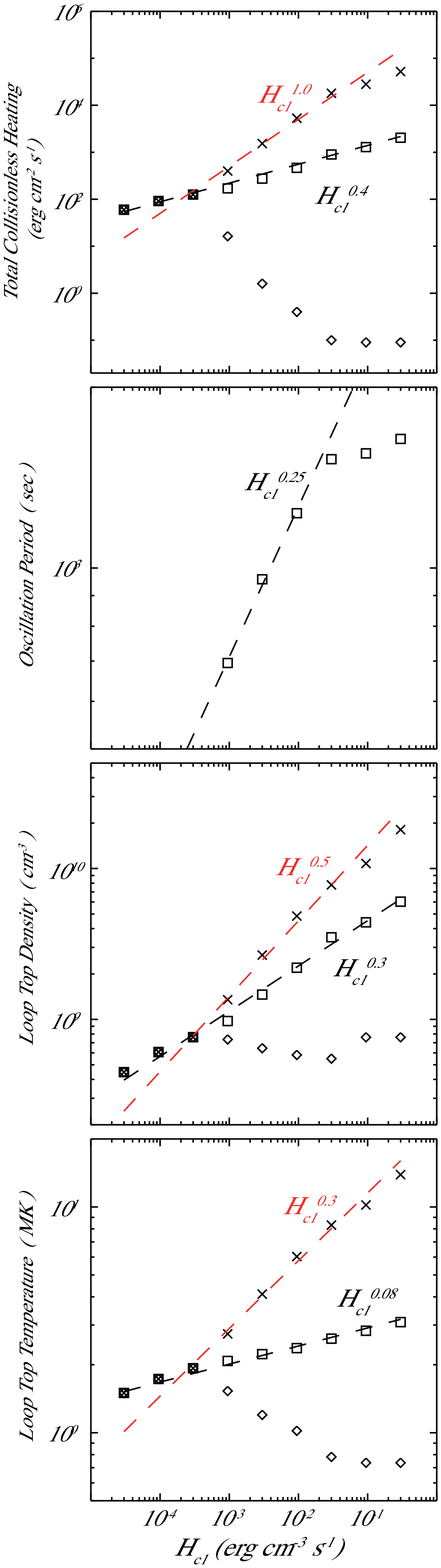}
\caption{ Characteristics of our model as a function of collisionless heating coefficient. Crosses, squares, and diamonds show the maximum, mean, and minimum values, respectively.}
\end{figure}

\end{document}